\documentclass[11pt]{article}
\input{psfig.sty}
\begin{document}

\title{
\textbf{Remarks about an ``exact'' RG theory of Goldstone modes}
}

\author{J. Kaupu\v{z}s
\thanks{E--mail: \texttt{kaupuzs@latnet.lv}} \\
Institute of Mathematics and Computer Science, University of Latvia\\
29 Rainja Boulevard, LV--1459 Riga, Latvia}

\date{\today}

\maketitle

\begin{abstract}
A renormalization group (RG) theory of Goldstone mode singularities
in the\linebreak $O(n>1)$--symmetric $\varphi^4$ model is discussed.
This perturbative RG theory is claimed to be asymptotically exact,
as regards the long--wave limit of the correlation functions,
where it predicts a purely Gaussian behavior of the transverse
correlation function. However, we show that the results of this
theory are incorrect, and the Gaussian behavior originates
from a rough error in mathematics. Other relevant perturbative
theories are discussed, as well.
\end{abstract}

\section{Introduction}

 Recently, our theory of grouping of Feynman diagrams~\cite{K1} has been
extended to include the region below the critical point~\cite{K3}.
The two--point correlation function in Fourier representation has
been studied, which depends on the wave vector ${\bf k}$.
Based on our method of systematic consideration of all diagrams,
we have shown in~\cite{K3} that the behavior of the longitudinal
($G_{\parallel}({\bf k})$) and transverse ($G_{\perp}({\bf k})$)
correlation functions in $O(n>1)$--symmetric $\varphi^4$ model below the
critical temperature $T_c$ is non--Gaussian. It means that
$G_{\perp}({\bf k}) \simeq a \, k^{-\lambda_{\perp}}$ and
$G_{\parallel}({\bf k}) \simeq b \, k^{-\lambda_{\parallel}}$
with exponents $d/2<\lambda_{\perp}<2$ and $\lambda_{\parallel}
=2 \lambda_{\perp} -d$ is the long--wave (at $k \to 0$) solution of our
equations at the spatial dimensionality $2<d<4$.
These results coincide also with a non--perturbative
renormalization group (RG) analysis provided in our paper.

  It turns out, however, that some perturbative RG theory~\cite{Law1,Law2},
which is claimed to be exact by its founders, contradicts our results --
it predicts a purely Gaussian behavior with $\lambda_{\perp}=2$.
We have put some effort to clarify this question and have found that
the Gaussian behavior in this theory originates from a rough 
error in mathematics.

\section{The critical analysis}

  We consider a $\varphi^4$ model with the Hamiltonian
\begin{equation} \label{eq:H}
H/T= \int \left( r_0 \varphi^2({\bf x}) + c (\nabla \varphi({\bf x}))^2 
+ u \varphi^4({\bf x}) - {\bf h} \varphi({\bf x}) \right) d {\bf x} \;,
\end{equation}
where the order parameter $\varphi({\bf x})$ is an
$n$--component vector with components $\varphi_i({\bf x})$, depending on the
coordinate ${\bf x}$, $T$ is the temperature, ${\bf h}$ is an external field.
The $\varphi^4$ model exhibits a nontrivial
behavior in close vicinity, as well as below the critical temperature
$T_c$, if the order parameter is an $n$--component vector
with $n>1$. The related long--wave divergence of the longitudinal
and transverse correlation functions (in Fourier representation)
at $T<T_c$ has been studied in~\cite{PaPo}
based on the hydrodynamical (Gaussian) approximation.
Essentially the same problem has been studied before in~\cite{Wagner}
in terms of the Gaussian spin--wave theory~\cite{Dyson}.
Later perturbative renormalization group (RG)
studies~\cite{FBJ,BW,WZ,Nel,BZ,Sch,Law1,Law2,Tu}
support the Gaussian approximation.
The RG method is claimed to be asymptotically exact.
Hence, we disprove this statement. Our analysis in~\cite{K3} predicts
a non--Gaussian behavior, and we show by general physical
arguments that it must be the true behavior to coincide
with the know rigorous results for the classical $XY$ model. 

According to the conventional
believ~\cite{FBJ,BW,WZ,Nel,BZ,Sch,Law1,Law2,Tu},
the transverse correlation function $G_{\perp}({\bf k})$
diverges like $k^{-\lambda_{\perp}}$ with $\lambda_{\perp}=2$ at
$k \to 0$ below $T_c$ for the systems
with $O(n>2)$ rotational symmetry. It corresponds to the
$G_{\parallel}({\bf k}) \sim k^{d-4}$ divergence of the
longitudinal correlation function. Besides, the singular structure
of the correlation functions is represented by an expansion
in powers of $k^{4-d}$ and $k^{d-2}$~\cite{Sch,Law1}.
Formally, our analysis in~\cite{K3} is consistent with these results at
$\lambda_{\perp}=2$, although in reality $\lambda_{\perp}<2$ holds.
Below we will show
that $\lambda_{\perp}<2$ holds near two dimensions at $n=2$.

As usually accepted in lattice models, here
we define that all the parameters of
the normalized Hamiltonian $H/T$~(\ref{eq:H}) are proportional
to the inverse temperature $1/T$. In this case $r_0$ is negative.
The assumption
that $G_{\perp}({\bf k}) \simeq a(T) \, k^{-2}$ holds
(with some temperature--dependent amplitude $a(T)$)
in the stable region below the critical point, i.~e., at
$T \le T_c/C$, where $C$ is an arbitrarily large constant,
leads to a conclusion that the critical temperature $T_c$
continuously tends to zero at $d \to 2$ (supposed $d>2$).
Really, at $\lambda_{\perp}=2$ we have
\begin{eqnarray} \label{eq:fufelis}
&&\left< \varphi^2({\bf x}) \right> = \frac{1}{V} \sum\limits_{i,{\bf k}}
G_i({\bf k}) \simeq M^2 \\
&&+ (2 \pi)^{-d} \left[ \int G'_{\parallel}({\bf k})
d{\bf k} + \frac{(n-1) \, S(d) \Lambda^{d-2} \, a(T)}{d-2} \right] \;,
\nonumber
\end{eqnarray}
where $G_i({\bf k})= \langle \mid \varphi_i({\bf k}) \mid^2 \rangle$
is the $i$--th component (one of which is
longitudinal, other ones -- transverse) of the correlation function,
$M$ is the magnetization, $\Lambda$
is the upper cutoff of the wave vector magnitude, and
$S(d)$ is the area of unit sphere in $d$ dimensions. Here
$G_{\parallel}({\bf k})
=G'_{\parallel}({\bf k})+\delta_{{\bf k},{\bf 0}} M^2 V$
is the Fourier transform of the real--space correlation function
$\langle \varphi_{\parallel}({\bf 0}) \varphi_{\parallel}({\bf x}) \rangle$,
whereas $G'_{\parallel}({\bf k})$ denotes the Fourier transform of
$\langle \varphi_{\parallel}({\bf 0}) \varphi_{\parallel}({\bf x})
\rangle - M^2$, where $\varphi_{\parallel}({\bf x})$ is the longitudinal
component of the order--parameter field.
Since the amplitude of the transverse fluctuations never can
vanish at a finite temperature, Eq.~(\ref{eq:fufelis}) implies that
the average $\left< \varphi^2({\bf x}) \right>$  diverges at
$T=T_c/C$ when $d \to 2$, if $T_c$ remains finite. Thus, we obtain an
unphysical result unless the critical temperature $T_c$
and, therefore, $a(T_c/C)$ tend to zero at $d \to 2$.

On the other hand, it is a rigorously stated fact~\cite{TKHT,FS} that 
the classical 2D $XY$ model undergoes the Kosterlitz--Thouless 
phase transition at a finite temperature $T_{\mbox{\tiny KT}}$.
It means that a certain structural order without the spontaneous
magnetization exists within the temperature region
$T < T_{\mbox{\tiny KT}}$. There is a
general tendency of disordering with decreasing the spatial
dimensionality $d$, and not vice versa.
Thus, since the structural order exists at $T < T_{\mbox{\tiny KT}}$
and $d=2$, some kind of order necessarily exists also
at $T < T_{\mbox{\tiny KT}}$ and $d>2$. Since the classical $XY$ model
undergoes the disorder $\rightarrow$ long--range order phase
transition at $d>2$, this obviously is the long--range order.
Thus, the critical temperature at
$d=2 + \varepsilon$ is $T_c \ge T_{\mbox{\tiny KT}} \ne 0$ for an
infinitesimal and positive $\varepsilon$, and it drops to zero by a
jump at $d=2 - \varepsilon$, as consistent with the rigorous
consideration in~\cite{TKHT}. 

The classical $XY$ model belongs to the same universality
class as the actual $\varphi^4$ model at $n=2$, which means that
both models become fully equivalent
after a renormalization (a suitable renormalization has been
discussed in~\cite{K3}). Thus,
$T_c$ does not vanish at $d \to 2$ (for $d>2$) also in the $\varphi^4$
model. In such a way, the assumption
$G_{\perp}({\bf k}) \simeq a(T) \, k^{-2}$ leads to a contradiction.
In the stable region $T<T_c/C$, the Gaussian approximation
$G_{\perp}({\bf k}) \simeq 1/(2ck^2)$
makes sense at finite not too small values of $k$. 
The above contradiction means that the Gaussian approximation
with $\lambda_{\perp}=2$ cannot be extended to $k \to 0$ in vicinity
of $d=2$. The contradiction is removed only if $\lambda_{\perp}<2$
holds at $d \to 2$ in the actual case of $n=2$.

It has been stated in~\cite{Law1,Tu} that the essentially Gaussian
result $\lambda_{\perp}=2$ of the perturbative RG theory should be
exact. However, the underlying method is not rigorous:
it is assumed without proof that the renormalized Hamiltonian
has the form of the Landau--Ginzburg--Wilson expansion with few terms
included, as in the original Hamiltonian~(\ref{eq:H}).
Besides the ``exact'' rusult is obtained by cutting
(at one loop order) a purely formal divergent perturbation series,
where the expansion parameters are in no sense small.
One claims~\cite{Law1,Law2,Tu} that the 
one--loop--diagram approximation provides asymptotically
exact results at $r_0 \to -\infty$ (which corresponds to $m_0 \to \infty$
or $m \to \infty$ in~\cite{Law1,Law2}) and, according to the provided there
renormalization group arguments, also in the long--wavelength limit
$k \to 0$. However, some of these ``exact'' results are rather unphysical. 
In particular, we find from Eq.~(3.6) in~\cite{Law2} and from
the formula $\langle \pi^2 \rangle=-6A/u_0$ given in the line
just below that
\begin{equation} \label{eq:mulkibas}
\langle \pi^2({\bf x}) \rangle = (N-1) \int \frac{d^dq}{q^2}
\end{equation}
holds, where $\pi({\bf x})$ is the transverse $(N-1)$--component
field. Eq.~(\ref{eq:mulkibas}) represents a senseless result,
since $\langle \pi^2({\bf x}) \rangle$
given by this equation diverges at $d \to 2$. It is clear that
$\langle \pi^2({\bf x}) \rangle$ cannot diverge in reality,
as it follows from the Hamiltonian density~(2.1)
in~\cite{Law2} (Hamiltonian~(\ref{eq:H}) in our paper): any field
configuration with diverging $\pi^2({\bf x})$
provides a divergent $\pi({\bf x})$--dependent term
\begin{equation} \label{eq:uh}
\sim  \frac{1}{2} \mid \nabla \pi({\bf x}) \mid^2
+ \frac{u_0}{4!} (\pi^2({\bf x}))^2
\end{equation}
in the Hamiltonian density and,
therefore, gives no essential contribution to the statistical averages.
The result~(\ref{eq:mulkibas}) corresponds to a poor approximation where
the second term in~(\ref{eq:uh}) is neglected.
In a surprising way, based on Ward identities,
authors of~\cite{Law2} and related papers have lost all the purely
transverse diagrams, generated by the term
$\frac{u_0}{4!} (\pi^2({\bf x}))^2$, and stated that this is the
exact result at $r_0 \to - \infty$.
These diagrams, evidently, cannot vanish due to $r_0 \to - \infty$,
simply, because they are independent of $r_0$. According
to~\cite{Law1,Law2}, the actual transverse term appears to be
hiden in a shifted longitudinal field $\bar s$, which 
is considered as an independent Gaussian variable
(cf.~Eqs.~(3.5) and~(3.6) in~\cite{Law1}). 
Obviously, this is the fatal trivial error which leads to the above
discussed unphysical result~(\ref{eq:mulkibas}), since $\bar s$
is not an independent field in reality -- the determinant
of the transformation Jacobian (from $\pi$, $s$ to $\pi$, $\bar s$)
is omitted in the relevant functional integrals!
Such an introduction of the shifted field to obtain
a formally Gaussian Hamiltonian does not make any sense,
because the transformation Jacobian then must be included and
the resulting model all the same is not Gaussian.
Since~(\ref{eq:mulkibas}) comes from
\begin{equation}
\langle \pi^2({\bf x}) \rangle = (N-1) \, (2 \pi)^{-d}
\int G_{\perp}({\bf k}) \, d^d k \;,
\end{equation}
the unphysical divergence of $\langle \pi^2({\bf x}) \rangle$
means that the predicted Gaussian form of the transverse correlation
function $G_{\perp}({\bf k})$ is incorrect.
Another aspect is that the method used
in~\cite{Law1,Law2} gives $\lambda_{\perp} \equiv 2$ also at $n=2$
in contradiction with our previous discussion concerning the known
rigorous results for the $XY$ model.

Our consideration does not contradict the conventional statement
(see~\cite{Tu} and references therein)
that the Gaussian spin--wave theory~\cite{Dyson} becomes exact at
low temperatures,
but only in the sense that it holds for any given nonzero
${\bf k}$ at $T \to 0$,
and in the limit $\lim\limits_{k \to 0} \lim\limits_{T \to 0}$
in particular. However, the actual limit of interest is $k \to 0$
or, equally, $\lim\limits_{T \to 0} \lim\limits_{k \to 0}$.
Therefore, it is impossible to make any rigorous conclusion  
regarding the exponent $\lambda_{\perp}$ (or any related exponent)
based on the fact that the Gaussian spin--wave theory becomes
exact at $T \to 0$.
One has to prove that the limits can be exchanged!
This problem persists also in the classical treatment of
the many--particle systems in~\cite{Wagner} which, in essence,
is based on the Gaussian spin wave theory at $T \to 0$.
This treatment, evidently, is not
exact, since it breaks down at $d \to 2$ for the
two--component ($n=2$) vector model (where $T_c$ remains finite 
and we fix the temperature $0<T<T_c$) just like we have
discussed already -- the average $\langle \pi^2({\bf x}) \rangle$
is given by the integral~(6.8) in~\cite{Wagner} which
diverges in this case (supposed
$(2 \pi)^{-3} d^3 k \to (2 \pi)^{-d} d^d k$).

A slightly different perturbative RG approach has been developed
in~\cite{BZ} to analyze the nonlinear $\sigma$ model. In this case
the modulus of $\varphi({\bf x})$ is fixed which automatically
removes the divergence of $\langle \pi^2({\bf x}) \rangle$.
A finite external field $h$ has been introduced there
to make an expansion. The correlation functions
have the power--like singularities of interest only at $h=+0$,
which means that in this case we have to consider the limit
$\lim\limits_{k \to 0} \lim\limits_{h \to 0}$, i.~e., the limit
$h \to 0$ must be taken first at a fixed nonzero $k$ ($p$ in formulae
used in~\cite{BZ}). The results in~\cite{BZ} are not rigorous since 
the expansions used there are purely formal, i.~e., 
they break down in this limit.
Besides, the renormalization of~\cite{BZ} is no more than
an uncontrolled approximation scheme:
contrary to the assumptions in~\cite{BZ}, it should be clear that the exact
renormalization is a fairly nontrivial problem which cannot be reduced
to a finding of only two renormalization constants.
If, e.~g., we make a real--space renormalization
of the Heisenberg model, say, with the scaling factor $s=2$,
then the statistically averaged block--spins of the Kadanoff transformation
(composed of $s^d$ original spins)
do not have a fixed modulus -- simply, the original model does not
include the constraint $\mid \varphi({\bf x}) \mid = const$
for the block averages.  It means that the transformation with any
finite $s$ yields a Hamiltonian of form different from the
original one, i.~e., the original Hamiltonian with merely renormalized
coupling constant can never be the fixed--point Hamiltonian.
 This is the common problem in the perturbative RG theory
that the renormalized Hamiltonian is assumed (or even ``proved''
like in~\cite{BZ}) to have the same form as the original one.
This, however, appears to be wrong in view of the exact renormalization.

In absence of rigorous proofs, theoretical predictions can be
supported by appropriate empirical data. However, the experimental
measurements of susceptibility $\chi$ depending on field $h$ in
isotropous ferromagnets like high--purity polycrystalline Ni~\cite{SRS}
are incompatible with the prediction
$\chi \sim h^{(d-4)/2}$ of the ``exact'' RG theory.
An explanation has been given in~\cite{Nel} that the divergence
of the susceptibility is not observed in real crystals due to
the symmetry breaking perturbations causing easy axes.
However, this argument is not valid for polycrystalline
materials, since they are practically isotropous
in a length scale which exceeds remarkably the size of grains.
Thus, the theory should work at $h \to 0$, where the relevant
length measure, i.~e. the correlation length, diverges.

\section{Conclusions}

The ``exact'' RG theory of Goldstone modes developed
in~\cite{Law1,Law2} and related papers is not exact, it is
unphysical and mathematically erroneous. The essential
claims about the Gaussian character of the $O(n>1)$--symmetric
$\varphi^4$ model below $T_c$ are based in~\cite{Law1}
on the fact that the Hamiltonian can be formally written
in an apparently Gaussian form. The author, however, forgot
to include the determinant of the transformation Jacobian
in the relevant functional integrals, according to which
the resulting model all the same is not Gaussian.
Surprisingly, the experts in the perturbative RG theory
have not mentioned this evidently rough error in their
theory till now. Other versions of the perturbative RG theory
do not have any rigorous mathematical basis, as well.
Besides, the prediction of the ``exact'' RG theory
that the susceptibility below $T_c$ diverges as $h^{-1/2}$
in three dimensions evidently disagree with the experimental
measurements in isotropous ferromagnets.

\end{document}